# Enhanced Battery Capacity Estimation in Data-Limited Scenarios through Swarm Learning

Jiawei Zhang
*Department of Electrical Engineering and Computer Science*
*University of Michigan*
Ann Arbor, MI 48109, USA
jiaweizh@umich.edu

Yu Zhang
*Department of Mechanical Engineering*
*National University of Singapore*
Singapore 11757, Singapore
wayne_zy@outlook.com

Wei Xu
*Department of Mechanical Engineering*
*National University of Singapore*
Singapore 11757, Singapore
wei.xu@u.nus.edu

Yifei Zhang
*Department of Mechanical Engineering*
*National University of Singapore*
Singapore 11757, Singapore
zhangyifei@u.nus.edu

Weiran Jiang
*Farasis Energy USA, Inc.*
Hayward, CA 94545, USA
wjiang@farasis.com

Qi Jiao
*Farasis Energy USA, Inc.*
Hayward, CA 94545, USA
qjiao@farasis.com

Yao Ren
*Farasis Energy USA, Inc.*
Hayward, CA 94545, USA
yren@farasis.com

Ziyou Song
*Department of Electrical Engineering and Computer Science*
*University of Michigan*
Ann Arbor, MI 48109, USA
ziyou@umich.edu

*Abstract*—Data-driven methods have shown potential in electric-vehicle battery management tasks such as capacity estimation, but their deployment is bottlenecked by poor performance in data-limited scenarios. Sharing battery data among algorithm developers can enable accurate and generalizable data-driven models. However, an effective battery management framework that simultaneously ensures data privacy and fault tolerance is still lacking. This paper proposes a swarm battery management system that unites a decentralized swarm learning (SL) framework and credibility weight-based model merging mechanism to enhance battery capacity estimation in data-limited scenarios while ensuring data privacy and security. The effectiveness of the SL framework is validated on a dataset comprising 66 commercial LiNiCoAlO2 cells cycled under various operating conditions. Specifically, the capacity estimation performance is validated in four cases, including data-balanced, volume-biased, feature-biased, and quality-biased scenarios. Our results show that SL can enhance the estimation accuracy in all data-limited cases and achieve a similar level of accuracy with central learning where large amounts of data are available.

## I. INTRODUCTION

Lithium-ion batteries are the dominant energy storage technology for electric vehicles (EVs), owing to their high energy density and long cycle life. To ensure the safe and reliable operation of EVs, robust battery monitoring and management systems are essential [1]. One critical function of such systems is battery capacity estimation, which helps track degradation and assess the remaining driving range of EVs [2]. However, accurately estimating battery capacity remains challenging due to the diverse aging patterns that arise under varying operating conditions.

Current battery capacity estimation approaches fall into two categories: model-based and data-driven. Model-based methods rely on physics-based electrochemical models and can provide accurate results through joint estimation of internal battery states. However, they are often complex and computationally intensive [3]. In contrast, data-driven approaches, which directly model the mapping relationship between features and target outputs, hold promises for enhanced estimation accuracy owing to their capability of handling complex data relationships and adaptability to diverse degradation mechanisms.

Recent studies have demonstrated the potential of data-driven techniques. Khaleghi et al. developed a data-driven Gaussian process regression estimator by extracting the time and frequency domain features during battery operation, achieving an estimation accuracy of about 1% [4]. Zhang et al. proposed a machine learning algorithm that extracts features from the battery charging voltage curves to estimate the battery capacity, and the accuracy can reach less than 1% [5]. To develop a realistic estimation method for practical EV usage scenarios, based on the relaxation voltage curve after battery charging, Zhu et al. extracted the statistical features of voltage relaxation curves and established an extreme gradient boost model, which can achieve an estimation error of 1.1% at various operating condition [2].

While these results are promising, the effectiveness of data-driven models depends heavily on the availability of large, high-quality datasets that capture a wide range of battery chemistries and use cases. In practice, such data are often fragmented and stored across different locations, creating a local learning (LL) environment where data and computing resources are decentralized [7]. This makes it difficult to train generalized models using isolated datasets. One promising solution is to share battery data among algorithm developers. For example, many battery researchers jointly proposed the concept of Battery Data Genome, to build a community of data hubs with standardized practices and enable flexible data sharing [6]. However, the development and deployment of data-sharing-based battery algorithms require an effective framework that can simultaneously preserve data privacy and ensure data security with fault-tolerance architecture, which is still lacking.

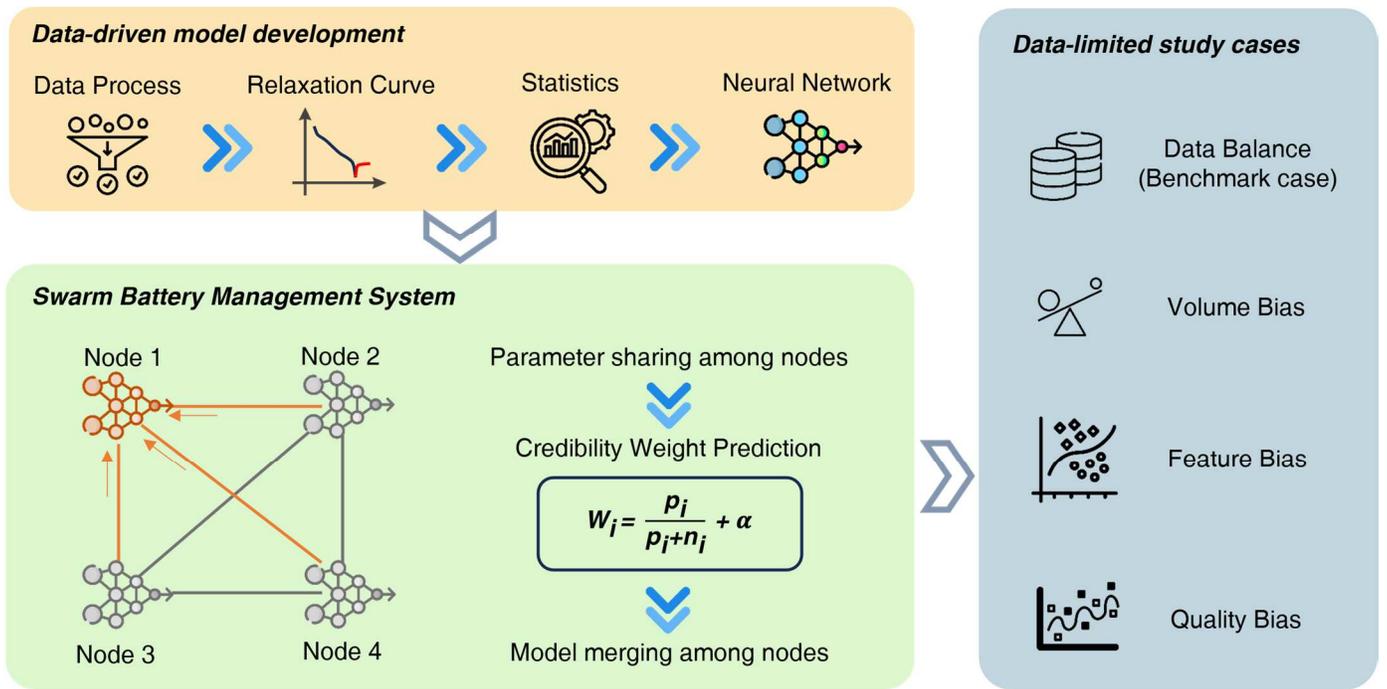

Fig. 1. Workflow of swarm learning-based battery capacity estimation.

Swarm learning (SL) [7], a decentralized machine learning approach, offers a potential solution. Originally applied in fields such as clinical prediction, SL enables secure, peer-to-peer model training without centralizing data. Unlike centralized learning (CL) [7], which aggregates both data and computation on a single server, or federated learning, which still depends on a central coordinator, SL uses blockchain technology to ensure secure communication and distributed coordination among participants. Each node in the network—representing an independent algorithm developer—trains models locally and shares only model parameters, preserving data privacy while improving collective performance. Despite its success in the medical field, the use of SL in battery management applications, including battery capacity estimation, has yet to be fully explored.

This paper aims to investigate the application of SL in enhancing the accuracy and generalizability of data-driven methods while paving the way for decentralized battery management strategies by addressing both the data security and privacy issues effectively. The overall workflow of development and validation of SL-based battery capacity estimation is shown in Fig. 1.

## II. SWARM BATTERY MANAGEMENT SYSTEM

### A. Swarm Learning Framework

In the proposed SL battery management framework, each node represents an independent algorithm developer working to build a battery capacity estimation model. Unlike traditional CL approaches, SL does not share raw training data across nodes. Instead, it aggregates the training outcomes—model parameters—through a decentralized process. The SL workflow consists of three main stages: initialization, iterative synchronization, and training. During initialization, each node independently sets up a machine-learning model and begins training on its local dataset. In the synchronization stage, nodes periodically share their model parameters, which are then merged into a global model through an iterative refinement process. The training continues until a predefined number of iterations is reached or until specific stopping criteria are met.

Specifically, A key component of the synchronization process is the credibility-weighted parameter merging, guided by a Credibility Weight Prediction Algorithm (CWPA) [8], illustrated in Fig. 2. In each synchronization round, both the local model of a node and the current global model are evaluated on a validation dataset. These evaluations produce credibility scores for each node (denoted as $C_i$) and for the global model (denoted as $C_a$). Based on these scores, a comparison is made: if $C_i$ is larger than $C_a$, the node's positive counter $p_i$ is increased; otherwise, the node's negative counter $n_i$ is increased. The final credibility weight for each node $W_i$ is computed as the ratio of the positive counter to the total number of evaluations, adjusted by a constant offset $\alpha$. These weights are then used to merge model parameters across all nodes using a weighted average, ensuring that more reliable models have a greater influence on the global model.

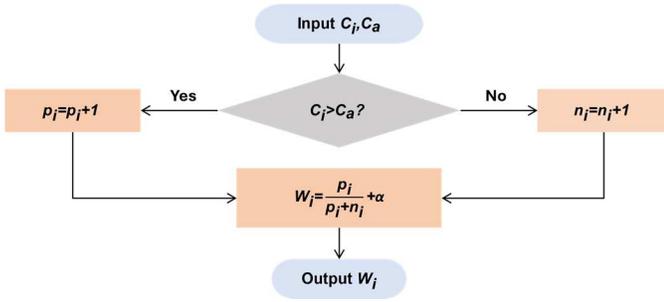

Fig. 2. Flowchart of credibility weight prediction algorithm

*B. Data-driven Model Development*

In this study, all nodes within the SL framework employ similar data-driven capacity estimators. The dataset and features used are based on a prior study [2], where the total dataset consists of 66 commercial LiNiCoAlO2 cells cycled under various operating conditions (i.e., constant-current constant-voltage operating mode with various C-rates and temperatures). The dataset is classified according to operating conditions as 'CYX-M/N', where 'X' represents the operating temperature of the battery, and 'M/N' represents the applied charge or discharge C-rate. For example, 'CY45-0.5/1' indicates a battery operating at 45°C with a charge C-rate of 0.5 and a discharge C-rate of 1.

To develop the data-driven models, three statistical features were extracted from the voltage relaxation curves: variance (Feature 1), skewness (Feature 2), and maximum value (Feature 3). Each data point in the dataset consists of these three features along with the corresponding battery capacity as the target label. This process resulted in over 20,000 data points available for training and evaluation. A feedforward neural network (FNN) is used as the capacity estimator across all SL nodes. The architecture includes three layers with neurons of 12, 8, and 1, respectively. The models were trained for 100 synchronization cycles, with one local training epoch per cycle and a credibility offset parameter $\alpha = 1$. To ensure robustness and generalization, 5-fold cross-validation was applied throughout the experiments. This experimental setup was designed to emulate practical conditions, including data heterogeneity caused by differences in fleet sizes, sensor precision, and operating environments. It provides a realistic and rigorous testbed for evaluating decentralized capacity estimation. The number of cells and data points associated with each operating condition is summarized in TABLE 1.

### III. EXPERIMENTAL VALIDATION AND DISCUSSION

*A. Case Study Setting*

In this work, four data-limited scenarios, including data-balanced, volume-biased, feature-biased, and quality-biased, are considered for validating the effectiveness of the SL framework with CWPA. The data-balanced case (TABLE 2) represents a benchmark where data distribution is well-balanced, and four nodes are assigned the same amount of data for the same working conditions. The volume-biased case (TABLE 3) represents the data-amount-limited scenario, such as uneven EV distribution across regions. The feature-biased case (TABLE 4) represents the data-diversity-limited scenario, where the ambient temperature is considered. The quality-biased case (TABLE 5) is designed to simulate the data-quality-limited scenarios with sensor measurement bias or data-storage fault. Specifically, Node 2 has 2000 data points at 25 ℃ (the actual temperature is 45 ℃) due to temperature sensor bias, while Node 3 has 1000 data points with wrongly labeled capacity values due to data-storage fault.

TABLE 1: DATASET AT VARIOUS OPERATING CONDITIONS

| Temperature (±0.2 °C) | Charge/discharge rate(C) | Number of cells | Number of data points |
|---|---|---|---|
|  | 0.25/1 | 7 | 1853 |
| 25 | 0.5/1 | 19 | 3278 |
|  | 1/1 | 9 | 260 |
| 45 | 0.5/1 | 28 | 15775 |

*B. Results and Discussions*

In the data-balanced case (Fig. 3), four nodes were designed, each assigned 2000 data points under the same working condition (CY45-0.5/1). The error rates for LL, SL, and CL were approximately 0.7%, 0.67%, and 0.64%, respectively. It can be observed that in the data-balanced case, LL, SL, and CL demonstrated comparable performance.

TABLE 2: DATA DISTRIBUTION IN THE DATA-BALANCED CASE

| Machine learning mode | Dataset | CY45-0.5/1 |
|---|---|---|
| LL/SL | Node1 | 2000 |
|  | Node2 | 2000 |
|  | Node3 | 2000 |
|  | Node4 | 2000 |
| CL | Train set | 8000 |
| Global | Validation set | 1000 |
| Global | Test set | 1000 |

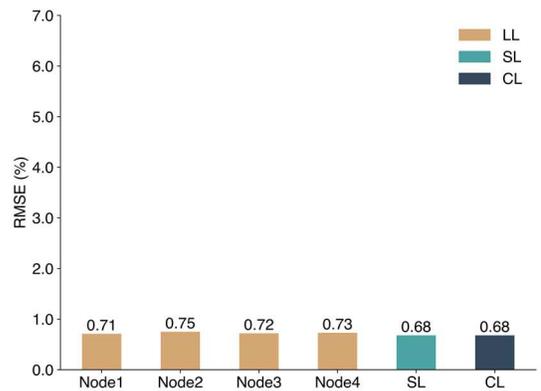

Fig. 3. Estimation results in data-balanced case.

In the volume-biased case (Fig. 4), three nodes were set up with 1000, 2000, and 5000 data points under the CY45-0.5/1 condition. LL exhibited poor performance in this scenario, with the node having the smallest dataset showing the highest error (3.0%), while the node with the largest dataset achieved an error of 1.5%, significantly lagging behind SL (0.76%) and CL (0.67%). When data bias exists, SL effectively balances data volume disparities, providing high-performance models for nodes with limited data and compensating for the shortcomings of LL. Although CL remains optimal, SL emerges as a superior distributed solution under privacy constraints.

TABLE 3: DATA DISTRIBUTION IN THE VOLUME-BIASED CASE

| Machine learning mode | Dataset | CY45-0.5/1 |
|---|---|---|
| LL/SL | Node1 | 1000 |
|  | Node2 | 2000 |
|  | Node3 | 5000 |
| CL | Train set | 8000 |
| Global | Validation set | 1000 |
| Global | Test set | 1000 |

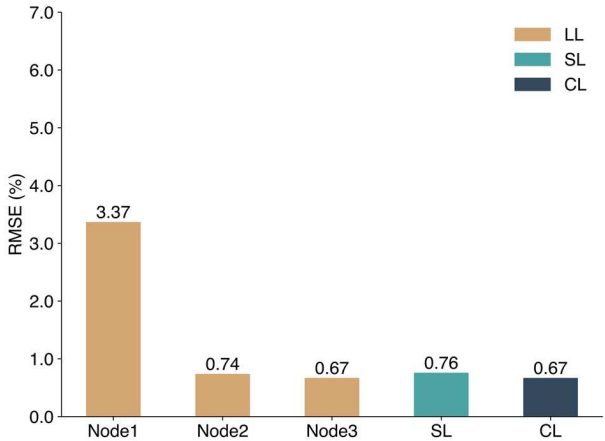

Fig. 4. Estimation results in volume-biased case.

In the feature-biased case (Fig. 5), data feature variations among nodes (e.g., different temperatures or charge/discharge rates) were simulated and categorized into absolute bias (single condition), strong bias (primary condition with minor others), and light bias (mixed conditions). Two nodes were assigned different proportions of high-temperature (CY45) and normal-temperature (CY25) data. LL achieved the highest error under absolute bias (3.49%) and the lowest under light bias (1.79%), while SL achieved errors of 2.22% and 1.70% under the same conditions, outperforming LL but still trailing CL (1.5%). It is evident that the greater the feature bias, the more significant the performance improvement from SL, such as the 1.3% reduction in error under absolute bias. Moreover, SL integrates multi-condition data, enhancing model generalization and approaching the performance of CL.

In the quality-biased case (Fig. 6), data quality variations among nodes were simulated, with three nodes configured: high-quality (CY45), low-quality (CY25), and artificially corrupted data (label tampering). The results showed that CL achieved errors of 0.73% and 6.16% for high-quality and low-quality nodes, respectively, while SL with CWPA achieved an error of only 1.11%, significantly outperforming SL without CWPA.

TABLE 4: DATA DISTRIBUTION IN THE FEATURE-BIASED CASE

| Machine learning mode | Dataset | Bias type | CY45-0.5/1 | CY25-0.5/1 |
|---|---|---|---|---|
| LL/SL | Node1 | Absolute | 0 | 2400 |
|  |  | Strong | 200 | 2200 |
|  |  | Light | 800 | 1600 |
|  | Node2 | Absolute | 2400 | 0 |
|  |  | Strong | 2200 | 200 |
|  |  | Light | 1600 | 800 |
| CL | Train set |  | 2400 | 2400 |
| Global | Validation set |  | 400 | 400 |
| Global | Test set |  | 400 | 400 |

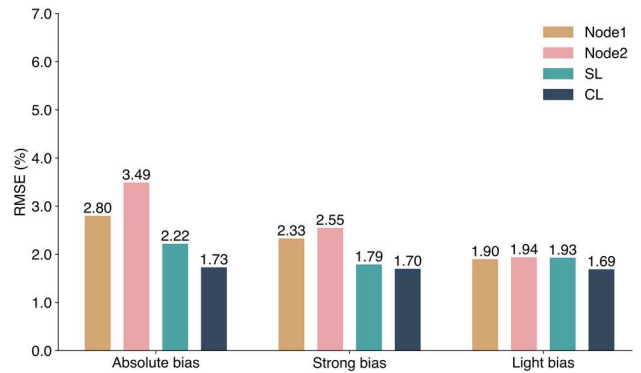

Fig. 5. Estimation results in feature-biased case.

TABLE 5: DATA DISTRIBUTION IN THE QUALITY-BIASED CASE

| Machine learning mode | Dataset | CY45-0.5/1 | CY25-0.5/1 |
|---|---|---|---|
| LL/SL | Node1 | 2000 | 0 |
|  | Node2 | 0 | 2000 |
|  | Node3 | 1000+1000 (modified) | 0 |
| CL | Train set | 3000+1000 (modified) | 2000 |
| Global | Validation set | 1000 | 0 |
| Global | Test set | 1000 | 0 |

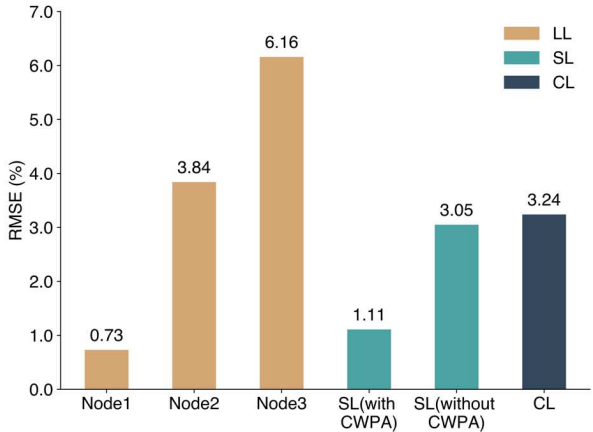
Fig. 6. Estimation results in quality-biased case.

## IV. Conclusion and Outlook

This study demonstrates that SL can effectively improve battery capacity estimation accuracy across various data-limited scenarios, outperforming traditional LL approaches. These results highlight the potential of a swarm battery management system (BMS) to enhance performance while maintaining data privacy and decentralization. Looking ahead, the methodology can be expanded, and additional applications of SL can be explored in future work. One promising direction is the integration of SL into EVs, where models could be trained using local BMS data—such as current and voltage measurements—for improved state estimation, such as the state of charge. This distributed approach would reduce dependence on large datasets from individual nodes and enhance estimation accuracy through collaborative learning. However, several challenges must be addressed to realize this vision.

First, the implementation of hardware infrastructures and platforms is critical for the successful application of swarm BMS. The blockchain and encryption technologies need to be combined with current algorithms to obtain the complete SL system to ensure secure and tamper-resistant data exchange. Second, the communication and computational limitations present technical hurdles. SL involves frequent synchronization among nodes, but EVs are often in motion, which can result in unstable network connections and delays. Furthermore, the limited processing power of in-vehicle BMS units may not support complex model training, necessitating lightweight algorithms and optimization techniques.

Moreover, ensuring consistent model aggregation in a dynamic, distributed environment is a complex task. Battery behavior and vehicle usage conditions change over time, requiring adaptive models that can evolve with new data. Designing SL frameworks capable of real-time adaptability remains an important area for research and development. Collaboration and trust among participating nodes also play a critical role. Some nodes may be unable or unwilling to contribute effectively, either due to limited resources or malicious intent. Implementing trust management systems is essential to ensure fairness, accountability, and resilience across the network.

In addition, regulatory and implementation challenges must be considered. Varying regional data protection laws complicate cross-border applications of SL, while the lack of industry-wide standards can lead to incompatible implementations. From a technical perspective, supporting SL may require upgrades to BMS hardware and increased maintenance efforts, contributing to higher deployment costs.

Despite these obstacles, ongoing advancements in blockchain, edge computing, and cloud-edge collaboration are expected to address many of these limitations. As these technologies mature, the development of swarm BMS holds promise to enable more reliable and efficient battery management strategies.